\documentclass[journal, twocolumn]{IEEEtran}

\usepackage{cite}
\usepackage{amsmath,amssymb,amsfonts}
\usepackage{gensymb}
\usepackage{float}
\usepackage{stfloats}
\usepackage{cancel}

\usepackage{array}
\usepackage{algorithmic}
\usepackage{graphicx}
\usepackage{textcomp}
\usepackage{xcolor}
\usepackage[hidelinks]{hyperref}
\usepackage{cancel}

\usepackage{amsthm}
\theoremstyle{definition}

\ifCLASSOPTIONcompsoc
    \usepackage[caption=false, font=normalsize, labelfont=sf, textfont=sf]{subfig}
\else
\usepackage[caption=false, font=footnotesize]{subfig}
\fi

\DeclareMathOperator{\sinc}{sinc}
\DeclareMathOperator{\rect}{rect}
\DeclareMathOperator{\tri}{tri}

\DeclareMathOperator{\sgn}{sgn}
\DeclareMathOperator{\Si}{Si}

\begin{document}

\title{Analysis and Approximation of Spatially Wideband Array Factor of Thinned Antenna Arrays}
 
\author{
    Marcin Wachowiak,~\IEEEmembership{Member,~IEEE,} 
    André Bourdoux,~\IEEEmembership{Senior~Member,~IEEE,}
    Sofie Pollin,~\IEEEmembership{Senior~Member,~IEEE,}%
    
    \thanks{
        Marcin Wachowiak and Sofie Pollin are with Interuniversitair Micro Electronica Centrum, 3001 Leuven, Belgium and also with the Katholieke Universiteit Leuven, 3000 Leuven, Belgium (e-mail: marcin.wachowiak@imec.be)}%
    \thanks{
        André Bourdoux is with Interuniversitair Micro-Electronica Centrum, 3001 Leuven, Belgium. (Corresponding author: \textit{Marcin Wachowiak})}
}

\maketitle

\begin{abstract}
This work investigates the spatially wideband (SWB) array factor (AF) of linear thinned antenna arrays.
The SWB AF is formulated as a matched filter and expressed as a spatially variant convolution of a spatially narrowband (SNB) AF and a SWB kernel. 
Originally derived for uniform arrays, the SWB AF approximation is revisited and extended to thinned arrays. The approximation is shown to remain accurate across a wide range of parameters.
Representing the SWB AF as a convolution offers an intuitive interpretation of how the bandwidth-aperture product modifies the SNB AF. Specifically, the SWB kernel acts as an averaging window for the SNB AF, suppressing the sidelobe levels (SL) of the SNB AF. 
In thinned arrays, the sidelobes away from the mainlobe can be treated as random, noise-like, making averaging and suppression particularly effective.
Next, the analytical formulas for the expected SL and peak-to-sidelobe level (PSL) in SNB thinned arrays are first introduced and subsequently extended to the SWB regime. 
Finally, simple closed-form expressions for the expected SL and PSL levels of the uniformly thinned arrays with a uniform spectrum are provided.
The expected SWB SL and PSL are formulated as functions of the bandwidth-aperture product and are shown to steadily decrease with distance from the mainlobe.

\end{abstract}

\begin{IEEEkeywords}
Antenna arrays, array factor, wideband, beamforming, beampattern, linear arrays, random arrays, nonuniform arrays, thinned arrays
\end{IEEEkeywords}

\section{Introduction}

\subsection{Problem Statement}

The antenna array factor (AF) is a key component that governs the spatial performance of wireless communication and sensing systems. The number of antennas and their spatial configuration determine the beampattern, angular resolution, achievable array gain, spatial selectivity and possible interference due to sidelobes \cite{van_trees_arrays, phased_arr_handbook, balanis_antenna}. 
The pursuit of improved performance in wireless communication and sensing systems motivates the adoption of increasingly larger apertures and wider bandwidths \cite{mimo_distributed_radar, 6g_large_arr_tutorial, gigantic_mimo}.
However, scaling up the aperture typically entails a linear or quadratic increase in the number of antenna elements and radio front-ends, leading to prohibitive costs in terms of hardware, power consumption and processing complexity.
To harness the benefits of large apertures while mitigating the costs, thinned antenna arrays should be considered \cite{amin_sparse, unconventional_arrays}.
Thinned antenna arrays offer a cost-effective solution to large-aperture implementation by increasing and randomizing the spacing between the antenna elements. The reduction in the number of antenna elements is achieved at the cost of increased sidelobe level (SL) \cite{lo_aperiodic, amin_sparse}.
With the same number of antenna elements, thinned arrays achieve a substantially larger aperture than uniformly spaced and fully populated arrays.
For a given bandwidth, the increase in the aperture size, achieved through thinning and nonuniform spacing, can violate the spatially narrowband (SNB) criterion and push the array into a spatially wideband (SWB) regime. In the SWB regime, the signal bandwidth and amplitude spectrum modify the conventional SNB AF, yielding a SWB AF.
Accurate system modeling requires quantifying and characterizing the effects of SWB operation on the AF of thinned antenna arrays.

\subsection{Relevant works}
The initial works on thinned antenna arrays were limited to the SNB regime.
In \cite{sidelobe_red_by_nonuniform_spacing, linear_arrays_var_spacing}, nonuniform element spacing is exploited to reduce the SL of linear arrays and achieve a narrow beamwidth with a lower number of antenna elements.
The space tapering and density-based thinning of planar and linear arrays are investigated in \cite{af_with_nonuniform_spacing, space_tapering_linear_and_planar, theory_of_unequally_spaced_arrays}.
The theory of unequally spaced arrays, enabling the design of desired radiation patterns, is introduced in \cite{theory_of_unequally_spaced_arrays, mathematical_theory_of_random_arrays}.
In \cite{skolnik_stat_density_tapered_arrays}, analytical SL formulas are provided for density-tapered thinned antenna arrays. It proposes using the amplitude taper of a fully populated array as a spatial density function, allowing to obtain the same beamwidth and SL in a thinned array as in a fully populated amplitude-tapered array.
The sidelobe distribution and peak SL based on probability in SNB random arrays are discussed in \cite{distribution_of_sidelobes_in_random_arrays}. The upper bounds on the height of the peak-to-sidelobe level (PSL) can be found in \cite{donvito_psl_of_random_arrays}.
Closed-form expressions for calculating the SL distributions and maximum PSL in random arrays are provided in \cite{sidelobe_distribution_arb_arr_distr, psl_gumbel_distribution}.
In \cite{steinberg_psl_of_random_arrays}, the PSL of random arrays with bandwidth is investigated. It incorporates the bandwidth in the PSL analysis; however, it does not provide closed-form expressions and detailed insights into how bandwidth suppresses the SL.
The antenna position and frequency diversity to suppress the sidelobes in random arrays is explored in \cite{sidelobe_reduction_element_and_freq_diversity}. 
In \cite{nanzer_spatial_filtering_grating_lobes}, the suppression of grating lobes with bandwidth is investigated. The work considers only two widely separated antennas in a mobile array scenario.
The requirements on bandwidth to suppress the grating lobes in ultra-wideband (UWB) systems are discussed in \cite{bw_requirement_for_suppresing_grating_lobes}. The paper focuses on a small array with a low number of antenna elements and large relative bandwidths.

\subsection{Contributions}

This work extends the concept of SWB AF to thinned antenna arrays. 
Formulating the SWB AF as a spatially variant convolution allows to separate the SWB contribution from the conventional SNB AF. 
Introducing the SWB kernel leads to an intuitive physical interpretation of how the large bandwidth-aperture product averages and smooths the resulting SWB AF.
Since the sidelobes of thinned arrays exhibit noise-like characteristics, they are particularly susceptible to averaging, making the suppression of the thinned array sidelobes by the SWB kernel particularly feasible.
The analytical expressions for expected SL and PSL of thinned arrays are extended to the SWB regime. The proposed SWB kernel approximations yield easy-to-use closed-form expressions of the SL and PSL as functions of the bandwidth-aperture product and the angle difference. 
Finally, the SL and PSL suppression in SWB thinned arrays is evaluated over a wide range of parameters, demonstrating the benefits of the SWB operation.

\section{Signal model}

\subsection{Spatially narrowband array factor (SNB AF)}
Consider an antenna array consisting of $M$ antennas indexed by $m$ forming a set $\mathcal{M}$. 
The array is designed at a center frequency $f_{\mathrm{c}}$ and the antennas are modeled as omnidirectional point radiators. Let vector $\mathbf{p}$ denote an arbitrary point of interest in space. The distance from the $m$-th antenna to the point $\mathbf{p}$ is denoted by $d_m(\mathbf{p})$. 
The beamforming position is given by $\mathbf{p}'$. The SNB AF is obtained by summing the weighted relative phase shift between $\mathbf{p}'$ and $\mathbf{p}$ across antenna elements as follows
\begin{align}
    \label{eq:nb_af}
    \mathrm{AF_{SNB}}(\mathbf{p}', \mathbf{p}, f) 
    &= \sum_{m \in \mathcal{M}} w(m)
    e^{-j2\pi \frac{f}{c} \Delta d_m(\mathbf{p}', \mathbf{p}) }.
\end{align}
where $w(m)$ is the amplitude weight (window) per antenna element and $\Delta d_m(\mathbf{p}', \mathbf{p}) = d_m(\mathbf{p}') - d_m(\mathbf{p})$ is the distance difference between $\mathbf{p}'$ and $\mathbf{p}$ for $m$-th antenna.

The SNB AF in \eqref{eq:nb_af} is equivalent to a weighted SNB matched filter and is valid at a single frequency $f$. The SNB AF can be considered accurate if the SNB condition is satisfied, that is, the propagation delay across the aperture is much smaller than the signal coherence time $\tau_D \ll \tau_B$. 
The maximum delay introduced by the antenna array is $\tau_D = D / c$, where $D$ is the array aperture and $c$ is the speed of light. The signal coherence time is given by $\tau_B = 1/B$, where $B$ is the signal bandwidth.
The SNB criterion is satisfied when $\frac{D B}{c} \ll 1$. Interesting to note that this criterion does not depend on the carrier frequency. It can be rewritten in terms of fractional bandwidth $B_{\mathrm{f}} = B / f_{\mathrm{c}}$ and aperture normalized with regard to wavelength,  $D_{\lambda} = D  / \lambda_{\mathrm{c}}$, as $D_{\lambda}B_{\mathrm{f}} \ll 1$, where the wavelength is $\lambda_{\mathrm{c}} = c / f_{\mathrm{c}}$.

Note that the SNB criterion very quickly becomes restrictive when considering large apertures or moderate fractional bandwidths. For example, for the SNB AF expression to remain accurate with $D_{\lambda} = 1000$, the fractional bandwidth must satisfy $B_{\mathrm{f}} \ll 0.1\%$. Alternatively, for a moderate fractional bandwidth of $B_{\mathrm{f}} = 10\%$, the total aperture must be $D_{\lambda} \ll 10$. 
As the aperture sizes and operating bandwidths increase, it becomes increasingly difficult to satisfy the SNB criterion and the SNB AF model becomes inaccurate. 
To accurately model the array performance for a large product of $D_{\lambda}B_{\mathrm{f}}$, the SWB AF must be accounted for. Typically, in the context of AF, the narrowband and wideband adjectives refer jointly to the spatial extent of the array and the signal bandwidth, not to the signal bandwidth alone. Here, to highlight the spatial and frequency aspects, the different operating regimes are referred to as spatially narrowband (SNB) and spatially wideband (SWB).

\subsection{Spatially wideband array factor (SWB AF)}
Consider that each antenna transmits or receives a signal of bandwidth $B$ centered at frequency $f_{\mathrm{c}}$. The signal is defined in the frequency domain as $S(f)$ with the total energy normalized to unity $\int_{-B/2}^{B/2} |S(f)|^2 \, df = 1$.
The SWB AF definition follows a matched-filter formulation, in line with the SNB AF. Due to non-negligible bandwidth, the matching is performed jointly in the spatial and frequency domains. The signal $S(f)$ is matched with its complex conjugate, arriving at $|S(f)|^2$. The SWB AF is obtained by integrating the SNB AF \eqref{eq:nb_af} over the signal bandwidth, weighted by the signal energy spectral density $|S(f)|^2$
\begin{align}
    \label{eq:wb_af}
    &\mathrm{AF_{SWB}}(\mathbf{p}', \mathbf{p}) 
    = \int_{-B/2}^{B/2} \left| S(f)\right|^2 
    \mathrm{AF_{SNB}}(\mathbf{p}', \mathbf{p}, f_{\mathrm{c}} + f) \, df.
\end{align}
Expanding \eqref{eq:wb_af} with \eqref{eq:nb_af} and separating the carrier frequency and the frequency variable components yields
\begin{align}
    \label{eq:wb_af_sep}
    \mathrm{AF_{SWB}}(\mathbf{p}', \mathbf{p}) 
    &= \sum_{m \in \mathcal{M}} w(m) e^{-j2\pi \frac{f_{\mathrm{c}}}{c} \Delta d_m(\mathbf{p}', \mathbf{p}) } \\
    & \quad \times \int_{-B/2}^{B/2} |S(f)|^2
    e^{-j2\pi \frac{f}{c} \Delta d_m(\mathbf{p}', \mathbf{p}) } \, df. \nonumber
\end{align}

This work discusses linear antenna array geometries as they represent the most fundamental and widely used configuration. 
In the far-field of a linear array, the point of interest $\mathbf{p}$ simplifies to the distance to the array weight center and angle with respect to the array normal $\mathbf{p} = [d, \theta]$.
The distance from $m$-th antenna to point $\mathbf{p}$ becomes
\begin{align}
    \label{eq:dist_per_ant}
    d_m(\mathbf{p})
    &= d - md_{\mathrm{a}} \sin{(\theta)}, 
\end{align}
where $d_{\mathrm{a}}$ is the antenna element spacing.
Matching to the transmitted or received signal in \eqref{eq:wb_af} introduces range dependency (resolution) in the SWB AF, as the waveform bandwidth provides temporal resolution.
To keep the conventional far-field interpretation of the AF and keep only the angular dependency of the AF, the points $\mathbf{p}$ and $\mathbf{p}'$ are fixed to the same distance $d'$. This corresponds to evaluating the angular AF at the beamforming distance (angular cross-cut) with respect to the center of the array.
The distance difference per antenna element, given that $d = d'$, reduces to
\begin{align}
    \label{eq:dist_diff_per_ant}
    \Delta d_m(\Delta u) 
    & = -m d_{\mathrm{a}} \Delta u,
\end{align}
where $\Delta u = u' - u = \sin{(\theta')} - \sin{(\theta)} $ is the direction sine difference.
After substituting \eqref{eq:dist_diff_per_ant} into \eqref{eq:wb_af_sep}, the SWB AF for the uniform linear antenna array (ULA) is
\begin{align}
    \label{eq:wb_af_ula}
    \mathrm{AF_{SWB}}(\Delta u) 
    &= \sum_{m \in \mathcal{M}} w(m) e^{j2\pi m \frac{d_{\mathrm{a}}}{\lambda_{\mathrm{c}}} \Delta u} \\
    & \quad \times \int_{-B/2}^{B/2} |S(f)|^2
    e^{j2\pi m \frac{f}{c} d_{\mathrm{a}} \Delta u} \, df. \nonumber
\end{align}
The inverse Fourier transform of $|S(f)|^2$ is, by definition, equivalent to the autocorrelation of the signal in the time domain as follows
\begin{align}
    \label{eq:autocorr}
    R(\tau) = \int_{-B/2}^{B/2} |S(f)|^2 e^{j2\pi f \tau} \, df.
\end{align}
Thus, the SWB AF from \eqref{eq:wb_af_ula} can be expressed as a product of the spatial window and the signal autocorrelation function from \eqref{eq:autocorr} as follows
\begin{align}
    \label{eq:wb_af_r_sum}
    \mathrm{AF_{SWB}}(\Delta u) 
    &= \sum_{m \in \mathcal{M}} w(m) R \left( \frac{m d_{\mathrm{a}}\Delta u }{c} \right) e^{j 2\pi m \frac{d_{\mathrm{a}}}{\lambda_{\mathrm{c}}} \Delta u }.
\end{align}
The SWB AF expression in \eqref{eq:wb_af_r_sum} does not provide straightforward insight into how the large bandwidth-aperture product modifies the SNB AF. More practical and intuitive expressions can be obtained by introducing suitable approximations.

\subsection{Approximation of spatially wideband array factor}
\label{sec:gen_wb_af_approx}

In this section, the SWB AF approximation from \cite{mw_wideband_af_approx} is briefly introduced and generalized to thinned antenna arrays. For a detailed derivation, refer to \cite{mw_wideband_af_approx}.
The SWB AF formulation in \eqref{eq:wb_af_r_sum} can be interpreted as a spatially variant convolution in which the signal autocorrelation term $R \left( \frac{m d_{\mathrm{a}}\Delta u }{c} \right)$ acts as a spatially varying weight across the array elements.
The product from \eqref{eq:wb_af_r_sum} can be expressed as a convolution using the discrete-time Fourier transform (DTFT) as follows
\begin{align}
    \label{eq:wb_af_dtft_conv}
    \mathrm{AF_{SWB}}(\Delta u)  
    = \Bigg( &\mathrm{DTFT} \left\{w(m)\right\} \\
    &\circledast
    \mathrm{DTFT}\left\{R \left( \frac{m d_{\mathrm{a}}\Delta u }{c} \right) \right\}
    \Bigg)
    \left( -\frac{d_{\mathrm{a}}}{\lambda_{\mathrm{c}}} \Delta u \right), \nonumber 
\end{align}
where $\circledast$ denotes periodic convolution. The negative sign in the argument of the convolution is due to an exponent sign difference in the definition of the DTFT and the SWB AF definition in \eqref{eq:wb_af_r_sum}.
The first factor in \eqref{eq:wb_af_dtft_conv} follows a similar formulation to the SNB AF
\begin{align}
    \label{eq:dtft_wm}
    \mathrm{DTFT} \left\{w(m)\right\} \left( v \right)  &= \sum_{m \in \mathcal{M}} w(m) e^{-j 2\pi m v} = W(v).
\end{align}
Evaluated for  $v = -\frac{d_{\mathrm{a}}}{\lambda_{\mathrm{c}}} \Delta u$ it becomes the SNB AF from \eqref{eq:nb_af} as follows $W\left( -\frac{d_{\mathrm{a}}}{\lambda_{\mathrm{c}}} \Delta u \right) = \mathrm{AF_{SNB}} ( \Delta u)$.
The second factor in \eqref{eq:wb_af_dtft_conv} is a SWB kernel defined as
\begin{align}
    \label{eq:dtft_wb_kern}
    \mathrm{DTFT} \left\{ R \left( \frac{m d_{\mathrm{a}}\Delta u }{c} \right) \right\}  \left(v\right)
    &= \sum_{m \in \mathcal{M}} R \left( \frac{m d_{\mathrm{a}}\Delta u }{c} \right) e^{-j 2\pi m v} \nonumber \\
    &= R_{\mathcal{M}}\left( \Delta u, v \right) 
\end{align}
Next, to obtain a closed-form expression of the SWB kernel in \eqref{eq:dtft_wb_kern}, the sum over $m$ is approximated by an integral using the Riemann sum approximation. The discrete sum over antenna elements is replaced by a continuous integral over the aperture. This approximation can be considered accurate for $B_{\mathrm{f}} \ll 1 / (2 d_{\mathrm{a}_{\lambda}})$, where $d_{\mathrm{a}_{\lambda}} = d_{\mathrm{a}} / \lambda$ is the antenna spacing normalized to wavelength
\begin{align}\
    \label{eq:cont_rd_approx}
    R_{\mathcal{M}}\left( \Delta u, v \right) \overset{B_{\mathrm{f}} \ll 1 / (2 d_{\mathrm{a}_{\lambda}}) }{\approx}  R_{D}\left( \Delta u, v \right).
\end{align}
After applying a change of variables, the continuous approximation of the SWB kernel can be expressed as an integral over the aperture as follows
\begin{align}
    \label{eq:rd_ap_int}
    R_{D}\left( \Delta u, v \right)
    &= \frac{1}{d_{\mathrm{a}}} \int_{-(D + d_{\mathrm{a}}) / 2}^{(D + d_{\mathrm{a}}) / 2}  R \left( \frac{ x \Delta u }{c} \right) e^{-j 2\pi \frac{x}{d_{\mathrm{a}}} v} \, dx. 
\end{align}
Approximating the discrete sum in \eqref{eq:dtft_wb_kern} as a continuous integral \eqref{eq:rd_ap_int} corresponds to substituting the DTFT with a continuous Fourier transform. Replacing the periodic output of the DTFT with the aperiodic continuous FT corresponds to approximating the periodic convolution in \eqref{eq:wb_af_dtft_conv} as aperiodic.
The aperiodic convolution accurately approximates the periodic one given that the SWB kernel decays before reaching the first alias position. The strict decay requirement is relaxed by the presence of the SNB AF in \eqref{eq:wb_af_dtft_conv} that acts as an anti-aliasing filter and suppresses the aliasing errors. For the alias contribution to be negligible, the mainlobe of the $\mathrm{AF_{SNB}}$ must be much narrower than the alias spacing $1 / D_{\lambda} \ll 1 / d_{\mathrm{a}_{\lambda}}$. Consequently, the approximation remains accurate even if the smoothness criterion in \eqref{eq:cont_rd_approx} is violated, provided the number of antennas is sufficiently large $M \gg 1$.
The approximation of the SWB AF from \eqref{eq:wb_af_dtft_conv} can be written as an aperiodic convolution as follows
\begin{align}
    \label{eq:wb_af_int_u_domain}
    &\mathrm{AF_{SWB}}(\Delta u) 
    \overset{M \gg 1}{\approx} \int_{-\infty}^{\infty} 
    \mathrm{AF_{SNB}}(u_{\tau}) \\
    & \qquad \qquad \qquad \qquad \times \tilde{R}_{D} \left( \Delta u, -\frac{d_{\mathrm{a}}}{\lambda_{\mathrm{c}}} \left( \Delta u - u_{\tau} \right) \right)\, du_{\tau}, \nonumber 
\end{align}
where $\tilde{R}_{D}(\Delta u, v) = \frac{d_{\mathrm{a}}}{\lambda_{\mathrm{c}}}  R_{D}(\Delta u, v)$ denotes the scaled continuous SWB kernel. 
By normalizing the aperture by wavelength, the scaled SWB kernel can be written as
\begin{align}
    \label{eq:rd_norm}
    &\tilde{R}_{D_{\lambda}} \left( \Delta u, -\frac{d_{\mathrm{a}}}{\lambda_{\mathrm{c}}} \left( \Delta u - u_{\tau} \right) \right) = \\
    &\qquad  = \int_{-(D_{\lambda} + d_{\mathrm{a}_{\lambda}}) / 2}^{(D_{\lambda} + d_{\mathrm{a}_{\lambda}}) / 2}  R \left( \frac{ x_{\lambda} \Delta u }{f_{\mathrm{c}}} \right) e^{j 2\pi x_{\lambda} \left( \Delta u - u_{\tau} \right)} \, dx_{\lambda}. \nonumber
\end{align}

As observed in \cite{mw_wideband_af_approx}, the amplitude tapering or windowing of the signal energy spectrum results in a more concentrated peak of the wideband kernel. A concentrated SWB kernel has a narrower width and so reduced sidelobe averaging and suppression capability.
Therefore, the following analyses are performed for a uniform amplitude spectrum, which corresponds to the best-case scenario offering the best sidelobe averaging performance. Moreover, the rectangular uniform spectrum is commonly found in OFDM or FMCW systems, making the analysis applicable to a wide range of systems.
For uniform spectrum $W(f) = 1/B$, $R(\tau) = \sinc{(B \tau)}$ the SWB kernel from \eqref{eq:rd_norm} simplifies to
\begin{align}
    \label{eq:rd_norm_uni_spec}
    &\tilde{R}_{D_{\lambda}} \left( \Delta u, -\frac{d_{\mathrm{a}}}{\lambda_{\mathrm{c}}} \left( \Delta u - u_{\tau} \right) \right) =  \\
    &= \frac{1}{B_{\mathrm{f}} \Delta u}
    \Bigg( 
    \Si{\left( (D_{\lambda} + d_{\mathrm{a}_{\lambda}}) \left( \frac{B_{\mathrm{f}}}{2}\Delta u + \Delta u - u_{\tau} \right) \right)} \nonumber  \\
    & \qquad \quad + \Si{\left( (D_{\lambda} + d_{\mathrm{a}_{\lambda}}) \left( \frac{B_{\mathrm{f}}}{2}\Delta u - \Delta u + u_{\tau} \right) \right)}
    \Bigg) \nonumber,
\end{align}
where $\Si{(x) = \int_{0}^{x} \sinc{(x)} \, dx} = \int_{0}^{x} \sin{(\pi x)} / (\pi x) \, dx$ is the normalized sine integral function.

\subsection{Thinned antenna arrays}

Consider a thinned antenna array obtained by removing the elements from a fully populated ULA. 
As previously defined, the total number of antenna elements in the fully populated array is $M$ and the corresponding index set is denoted by $\mathcal{M}$.
The thinning of the antenna array is performed based on density tapering, in which the probability that the antenna slot is occupied varies across the aperture and is determined by the spatial density profile.
The density tapering allows designing the sidelobe behaviour of the thinned arrays based on an analogy to the amplitude tapering of fully populated arrays \cite{skolnik_stat_density_tapered_arrays}.
Consider a function $f(m)$, which serves as a basis for defining the spatial density of antenna elements across the array. 
The function is normalized so that the maximum value is unity $\max{\{ f(m) \}} = 1$.
The total number of antennas in the thinned array is controlled by the fill parameter $\eta \in(0,1]$ which determines the expected number of antennas as a fraction of the number of antennas in the fully populated array. The number of antennas in the thinned array, denoted by $M_{\mathrm{th}}$, becomes a random variable and on average is equal to
\begin{align}
    \label{eq:exp_n_ant}
    \mathbb{E}[M_{\mathrm{th}}] = \overline{M_{\mathrm{th}}} = \eta M.
\end{align}

The thinning of the array is done based on the binary occupation variable $B_m \in \{0, 1\}$, where unity denotes that the antenna element is present; otherwise, the slot is empty.
The probability that the $m$-th slot is occupied is given by the Bernoulli distribution as follows $\mathrm{Pr}\{B_m = 1\} = p_m$ and $\mathrm{Pr}\{B_m = 0\} = 1 - p_m$. The $p_m$ is the activation probability defined as the scaled spatial density profile $p_m = \alpha |f(m)|$, where $\alpha$ is an additional scaling parameter that maintains a desired number of antenna elements in the thinned array and is calculated based on $\eta$. The absolute value is used to account for possible negative values of the spatial density profile.
The expected number of antenna elements can be written using activation probability as
\begin{align}
    \label{eq:exp_n_ant_pm}
    \mathbb{E}\left[M_{\mathrm{th}} \right] 
    = \mathbb{E}\left[ \sum_{m \in \mathcal{M}} B_m \right] 
    &= \sum_{m \in \mathcal{M}} p_m 
\end{align}
Expanding with the definition of $p_m$ and solving for $\alpha$ gives
\begin{align}
    \alpha &= \frac{ \eta M}{\sum_{m \in \mathcal{M}} |f(m)|}
\end{align}
Since $p_m$ is probability, it must satisfy $0 \leq  p_m \leq 1$, Hence $\alpha |f(m)| \leq 1$ leading to constraint on the maxium value of fill parameter $\eta \leq \frac{\sum_{m \in \mathcal{M}} |f(m)|}{M \max_m{|f(m)|}}$.
As the number of antennas grows large, the maximum $\eta$ becomes solely a function of the spatial density profile. The maximum value of $\eta$ for large arrays $M \to \infty$ can be obtained by approximating the sum in the inequality as an integral, yielding $\eta \leq \int_{0}^{1}{f(x)} \, dx$. Table \ref{tab:eta_max_per_wdw} presents the maximum achievable value of $\eta$ for different spatial density functions.
\begin{table}[tb]
    \caption{$\eta_{\mathrm{max}}$ per spatial density profile}
    \label{tab:eta_max_per_wdw}
    \centering
    \begin{tabular}{
      >{\centering\arraybackslash}m{0.15\linewidth}<{}
      |>{\centering\arraybackslash}m{0.15\linewidth}<{}
      |>{\centering\arraybackslash}m{0.15\linewidth}<{}
      |>{\centering\arraybackslash}m{0.15\linewidth}<{}
      |>{\centering\arraybackslash}m{0.15\linewidth}<{}
    }
         $f(m)$ & Uniform & Hamming & Hann & Blackman \\
         \hline 
         $\max{\{\eta\}}$& 1.0 & 0.54 & 0.5 & 0.42 \\ 
    \end{tabular}
\end{table}

\subsection{Generalization of the SWB AF approximation for thinned arrays}
The universal SWB AF approximation introduced in Sec. \ref{sec:gen_wb_af_approx} was derived for uniform antenna element spacing. Nevertheless, it is also applicable to thinned antenna arrays without requiring any substantial modifications to the array signal model. 
The universality and elegance of the SWB AF approximation lies in the separation of the SWB AF into the individual contributions from SNB AF and the SWB kernel. 

By recognizing that the spatial window $w(m)$ in \eqref{eq:wb_af_r_sum} and \eqref{eq:dtft_wm} can be an arbitrary function, the array signal model is extended to arbitrary AFs. Note that the density taper function $f(m)$ is a real function and can achieve negative values, which correspond to phase-flipped elements. 
For some realization of $b_m$ with $M_{\mathrm{th}}$ active antennas, the spatial window $w(m)$ becomes a scaled bipolar variable as follows 
\begin{align}
    \label{eq:norm_spatial_wdw}
    w(m) &= \frac{ b_m  \sgn{(f(m))}}{ \sqrt{M_{\mathrm{th}}}}.
\end{align}
The sign function of $f(m)$ conserves the polarity and phase of elements determined by the spatial density profile $f(m)$.
The scaling is introduced to ensure the precoder power is normalized to unity, $\sum_{m \in M} |w(m)|^2 = 1$.

\begin{figure}[b]
    \centering
    \includegraphics[width=\linewidth]{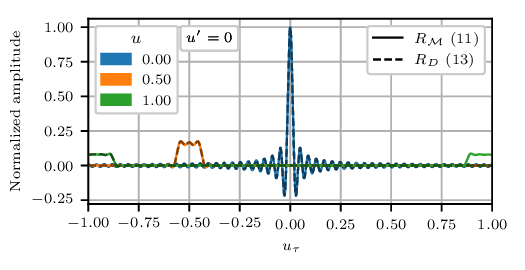}
    \caption{Discrete vs continuous SWB kernel for uniform spectrum, $u'=0$, $D_{\lambda}=50$, $d_{\mathrm{a}_{\lambda}} = 0.5$ and $B_{\mathrm{f}} = 0.25$.}
    \label{fig:rd_vs_approx}
\end{figure}
To guarantee the accuracy of the SWB AF approximation, the approximation criteria in \eqref{eq:cont_rd_approx} and \eqref{eq:wb_af_int_u_domain} are revisited and discussed in the context of the thinned arrays.
The smoothness criterion from the continuous approximation in \eqref{eq:cont_rd_approx} remains unchanged, as the SWB kernel in \eqref{eq:dtft_wb_kern} is evaluated over the full $\mathcal{M}$ grid irrespective of the occupancy given by $w(m)$. 
Thus, the continuous approximation condition in \eqref{eq:cont_rd_approx} given by $B_{\mathrm{f}} \ll 1 / (2 d_{\mathrm{a}_{\lambda}})$ is the same for thinned antenna arrays. 
Fig. \ref{fig:rd_vs_approx} compares the discrete SWB kernel with its continuous approximation for a uniform spectrum. As discussed before, the continuous approximation is aperiodic, while the discrete kernel exhibits aliases at $1 / d_{\mathrm{a}_{\lambda}}$. For a spatial grid that satisfies the Nyquist sampling $d_{\mathrm{a}_{\lambda}}$, for the aliasing error to be substantial would require a very large fractional bandwidth and would occur only for large $\Delta u$.

\begin{figure}[tb]
    \centering
    \includegraphics[width=\linewidth]{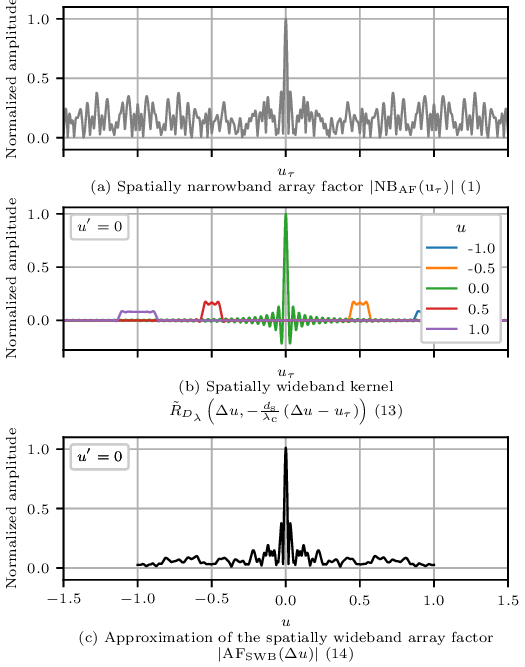}
    \caption{Illustration of spatially variant convolution and its components yielding the approximation of the SWB AF for a uniformly thinned array with $\eta = 0.25$, uniform spectrum window, $u'=0$, $D_{\lambda}=50$, $d_{\mathrm{a}_{\lambda}} = 0.5$ and $B_{\mathrm{f}} = 0.25$.}
    \label{fig:conv_visualization}
\end{figure}

The second approximation criterion from \eqref{eq:wb_af_int_u_domain} also remains unchanged, as the mainlobe width and alias spacing for thinned arrays with element grid spacing $d_{\mathrm{a}_{\lambda}}$ are the same as for fully populated arrays, given that the total aperture of the array is conserved. The approximation remains accurate for $1 / D_{\lambda} \ll 1 / d_{\mathrm{a}_{\lambda}}$, which for thinned arrays becomes $ M_{\mathrm{th}} \gg 1$. The approximation remains accurate as long as the number of antenna elements in the thinned array is sufficiently large.
Similarly, as for fully populated arrays, the criterion on the smoothness is relaxed by the presence of SNB AF, which acts as a narrow spatial filter reducing the approximation error due to aliasing. 
Fig. \ref{fig:conv_visualization} illustrates the individual components constituting the SWB AF approximation for a uniformly thinned array.
Fig. \ref{fig:conv_visualization}a shows the SNB AF of a thinned array given by \eqref{eq:nb_af}. The SNB AF of a thinned array exhibits high noise-like sidelobes due to thinning. 
Fig. \ref{fig:conv_visualization}b illustrates the SWB kernel for a uniform spectrum from \eqref{eq:rd_norm_uni_spec} as a function of $u_{\tau}$ for a few selected $u$ values. 
Finally, Fig. \ref{fig:conv_visualization}c shows the resulting SWB AF of the thinned array with bandwidth, which is obtained by integrating the product of the SNB AF and the SWB kernel across $u_{\tau}$ for each $u$ according to \eqref{eq:wb_af_int_u_domain}. The SLs of the SWB AF in Fig. \ref{fig:conv_visualization}c are noticeably lower compared to those of the SNB AF in Fig. \ref{fig:conv_visualization}a, thanks to the averaging effect introduced by the SWB kernel. Integrating the noise-like sidelobes of a thinned array by the sliding SWB kernel substantially reduces their power by averaging them across the $u$ domain. 

\begin{figure}[tb]
    \centering
    \includegraphics[width=\linewidth]{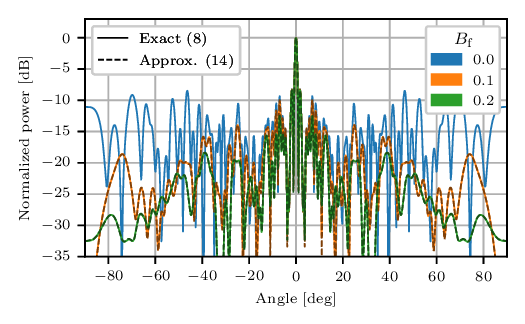}
    \caption{Comparison between the exact SWB AF and the approximation for a uniformly thinned array $\eta=0.25$, $D_{\lambda}=50$, $d_{\mathrm{a}_{\lambda}} = 0.5$, uniform spectrum window, $u'=0$ and selected fractional bandwidths.}
    \label{fig:wb_af_approx_vs_fbw}
\end{figure}

\begin{figure}[tb]
    \centering
    \includegraphics[width=\linewidth]{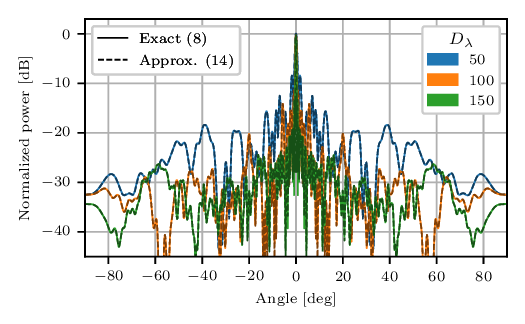}
    \caption{Comparison between the exact SWB AF and the approximation for a uniformly thinned array $\eta=0.25$, $d_{\mathrm{a}_{\lambda}} = 0.5$, $B_{\mathrm{f}} = 0.2$, uniform spectrum window, $u'=0$ and selected aperture sizes.}
    \label{fig:wb_af_approx_vs_dap}
\end{figure}

Fig. \ref{fig:wb_af_approx_vs_fbw} illustrates the SWB AF of a uniformly thinned array for a few values of the fractional bandwidth. The increase in $B_{\mathrm{f}}$ widens the SWB kernel that acts as an averaging window for the sidelobes, effectively reducing the SL. The SL reduction improves with angle as the SWB kernel widens with increasing $\Delta u$, as shown in Fig. \ref{fig:rd_vs_approx}. 
Fig. \ref{fig:wb_af_approx_vs_dap} illustrates the SWB AF of a uniformly thinned array for a fixed fractional bandwidth and a few selected values of aperture sizes. Notably, increasing the aperture does not widen the SWB kernel but increases the variability of the sidelobes in $\Delta u$, allowing the SWB kernel of the same width to capture a larger number of randomly varying sidelobes, averaging and suppressing them more effectively.


Note that the SWB AF approximation in Sec. \ref{sec:gen_wb_af_approx}, can be extended to arbitrarily spaced nonuniform antenna arrays (not restricted to the predefined position grid) by fixing $D_{\lambda} = (M-1) d_{\mathrm{a}_{\lambda}}$ and letting the antenna position grid becomes increasingly dense by $d_{\mathrm{a}_{\lambda}} \to 0$ and $M \to \infty$. Under this limit the position of the $m$-th antenna on the X axis in the thinned array becomes a continuous variable denoted by $x_{m}$.
Under the infinitely dense sampling limit, the AF from \eqref{eq:nb_af} and \eqref{eq:dtft_wm} for nonuniform array can be written as
\begin{align}
    \label{eq:nb_af_th}
    \mathrm{AF_{SNB}} ( \Delta u)
    &= \sum_{m \in \mathcal{M}} w(x_m)  e^{j2\pi \frac{x_{m}}{\lambda_{\mathrm{c}}} \Delta u} \nonumber \\
    &= \frac{1}{\sqrt{M_{\mathrm{th}}}} \sum_{m \in \mathcal{M}_{\mathrm{th}}} e^{j2\pi \frac{x_{m}}{\lambda_{\mathrm{c}}} \Delta u}.
\end{align}
In the infinitely dense sampling limit, the continuous SWB kernel from \eqref{eq:rd_norm} becomes
\begin{align}
    \label{eq:rd_norm_lim}
    &\tilde{R}_{D_{\lambda}} \left( \Delta u, -\left( \Delta u - u_{\tau} \right) \right) = \\
    & \qquad\qquad\qquad   = \int_{-D_{\lambda} / 2}^{D_{\lambda}  / 2}  R \left( \frac{ x_{\lambda} \Delta u }{f_{\mathrm{c}}} \right) e^{j 2\pi x_{\lambda} \left( \Delta u - u_{\tau} \right)} \, dx_{\lambda}. \nonumber
\end{align}
By plugging in \eqref{eq:nb_af_th} and \eqref{eq:rd_norm_lim} into \eqref{eq:wb_af_int_u_domain} the SWB AF approximation for arbitrary nonuniform arrays is
\begin{align}
    \label{eq:wb_af_int_u_domain_dense_lim}
    &\mathrm{AF_{SWB}}(\Delta u) 
    \overset{M_{\mathrm{th}} \gg 1}{\approx} \int_{-\infty}^{\infty} 
    \mathrm{AF_{SNB}}(u_{\tau}) \\
    & \qquad\qquad\qquad\qquad\qquad\quad  \times \tilde{R}_{D_{\lambda}} \left( \Delta u, -\left( \Delta u - u_{\tau} \right) \right)\, du_{\tau}. \nonumber 
\end{align}

\subsection{Statistical sidelobe level (SL) analysis of SNB AF}
\label{sec:sl_analysis}
In this section, the statistical model of the SL in SNB density-tapered thinned arrays is introduced from \cite{skolnik_stat_density_tapered_arrays}.
To obtain analytical formulas quantifying the SL in thinned arrays, consider a linear antenna array with arbitrarily located antenna elements along the X axis. The position of $m$-th antenna element is given by a continuous variable $x_m$ and the spatial density profile becomes a continuous function $f(x_m)$. 
The analysis here is conducted with respect to SNB AFs and, for compactness of notation, the $\mathrm{SNB}$ subscript in the $\mathrm{AF}$ is omitted.
In the following, each summation over the antenna elements is taken over the set $\mathcal{M}$. For compactness and clarity of the notation, in sums only the index $m$ is used. 
For an array with randomly positioned elements, the spatial window is a random variable defined as follows
\begin{align}
    \label{eq:spatial_window_rand}
    W(x_m) = \frac{B_m \sgn{(f(x_m))}} {\sqrt{\overline{M_{\mathrm{th}}}}}
\end{align}
The scaling by the expected number of antenna elements from \eqref{eq:exp_n_ant} is introduced to keep the total expected power of the spatial window normalized to unity $\mathbb{E} \left[ \sum_{m} |W(x_m)|^2\right] = 1$.
The expected value of a random spatial window is equal to the deterministic spatial window, as follows
\begin{align}
    \label{eq:determ_spatial_wdw}
    w(x_m) = \mathbb{E} \left[ W(x_m) \right] 
    &= \frac{ \alpha f(x_m) }{\sqrt{ \overline{M_{\mathrm{th}}}}} .
\end{align}
Given the spatial window in \eqref{eq:spatial_window_rand}, the expected value of the AF of a thinned array can be written as
\begin{align}
    \label{eq:exp_ampl_wdw_array}
    \mathbb{E} \left[ \mathrm{AF_{th}}(\Delta u) \right]
    &= \mathbb{E} \left[ \sum_{m} W(x_m) e^{-j 2 \pi \frac{x_m}{\lambda_{\mathrm{c}}} \Delta u}  \right] \nonumber \\
    &= \sum_{m}  w(x_m)  e^{-j 2 \pi \frac{x_m}{\lambda_{\mathrm{c}}} \Delta u} \nonumber \\
    &= \mathrm{AF_{ wdw}} (\Delta u).
\end{align}
The expected value of the AF of a density-tapered thinned array corresponds to the AF of an equivalent deterministic amplitude-tapered \eqref{eq:determ_spatial_wdw} array denoted by $\mathrm{AF_{wdw}} (\Delta u)$. 

The power of a deterministic amplitude-tapered AF can be obtained by writing it as a sum of diagonal and off-diagonal components.
Since real spatial profiles $f(x_m)$ are considered allowing to write $w(x_m)^* = w(x_m)$, the power of the AF of amplitude-tapered array simplifies to
\begin{align}
    \label{eq:af_ampl_wdw}
    &|\mathrm{AF_{wdw}} (\Delta u)|^2 = 
    \sum_{m} w(x_m)^2  + \\
    &\qquad\qquad\qquad + \sum_{\substack {m  \\ m\neq n}} \sum_{n} w(x_m) w(x_n)  e^{-j 2 \pi \frac{(x_m - x_n)}{\lambda_{\mathrm{c}}} \Delta u}. \nonumber
\end{align}
The expected power of the AF of a random thinned array is obtained similarly to \eqref{eq:af_ampl_wdw} by taking the expectation as follows
\begin{align}
    \label{eq:expected_th_af_pwr}
    & \mathbb{E} \left[ |\mathrm{AF_{th}}(\Delta u)|^2 \right]
    =\mathbb{E} \left[ \sum_{m} W(x_m)^2 \right] \\
    & \qquad\qquad + \mathbb{E} \left[ \sum_{\substack {m  \\ m\neq n}}\sum_{n}  W(x_m) W(x_n)  e^{-j 2 \pi \frac{(x_m - x_n)}{\lambda_{\mathrm{c}}} \Delta u} \right]. \nonumber
\end{align}
The first term in \eqref{eq:expected_th_af_pwr} reduces to unity due to expected power normalization from \eqref{eq:spatial_window_rand}.
The second component, the double sum in \eqref{eq:expected_th_af_pwr}, which is the expectation of the product, can be written in terms of the product of the expectations based on the observation that $W(x_m)$ and $W(x_n)$ are drawn independently for $m \neq n$, yielding
\begin{align}
    \label{eq:prod_of_expectations}
    &\sum_{\substack {m  \\ m\neq n}}\sum_{n}  \mathbb{E} \left[ W(x_m) \right] 
    \mathbb{E} \left[ W(x_n) \right] e^{-j 2 \pi \frac{(x_m - x_n)}{\lambda_{\mathrm{c}}} \Delta u} \nonumber \\
    &= \sum_{\substack {m  \\ m\neq n}} \sum_{n} w(x_m) w(x_n)e^{-j 2 \pi \frac{(x_m - x_n)}{\lambda_{\mathrm{c}}} \Delta u}.
\end{align}
An alternative formulation of the expected power of the thinned antenna AF is obtained by recognizing that \eqref{eq:prod_of_expectations} in \eqref{eq:expected_th_af_pwr} can be expressed in terms of a deterministic amplitude-tapered AF from \eqref{eq:af_ampl_wdw} as follows
\begin{align}
    \label{eq:af_thinned_decoupled}
    &\mathbb{E} \left[ |\mathrm{AF_{th}}(\Delta u)|^2 \right]
    = |\mathrm{AF_{wdw}}(\Delta u)|^2 + \underbrace{ \left(1 - \sum_m w(x_m)^2\right)}_{ \sigma^2 },
\end{align}
The second term in \eqref{eq:af_thinned_decoupled} is the mean sidelobe power, equal to the variance of the active number of antenna elements normalized by their expected number.
This significant result from \cite{skolnik_stat_density_tapered_arrays} enables expressing the expected AF of a thinned array as a sum of a deterministic amplitude-tapered AF and a noise-like sidelobe term $\sigma^2 = |Z(\Delta u)|^2$. Note that the fluctuation (sidelobes) term is uniform across $u$. 
To calculate the average SL, the sidelobes from the deterministic component of the AF are considered negligible compared to the statistical part in \eqref{eq:af_thinned_decoupled}. This approximation is valid in the far-sidelobe region - sufficiently far from the mainlobe and a few of its first sidelobes ($u$ sufficiently away from the mainlobe $u'$).
The average SL is obtained as a ratio of the average sidelobe power to the peak power of the deterministic AF as
\begin{align}
    \label{eq:exp_avg_sl}
    \mathbb{E} \left[ \mathrm{SL} \right]  
    = \overline{\mathrm{SL}} 
    &= \frac{\sigma^2 } {\max_{\Delta u}{ \left\{ \mathrm{|AF_{wdw}}(\Delta u) |^2 \right\}}}.
\end{align}

For even and non-negative spatial profile $f(x_m)$, the maximum power of the deterministic AF occurs for $\Delta u = 0 $ where $|\mathrm{AF_{wdw}}(0)|^2 = \overline{M_{\mathrm{th}}}$. The average SL then becomes
\begin{align}
    \label{eq:exp_avg_sl_even_nonneg}
    \overline{\mathrm{SL}}
    = \frac{\sigma^2}{\overline{M_{\mathrm{th}}}}
    &= \frac{\sum_m p_m (1 - p_m)}{  \overline{M_{\mathrm{th}}}^2}.
\end{align}
For uniform spatial profile $p_m = \eta$ it simplifies to a well known result $\overline{\mathrm{SL}}_{\mathrm{uni}} = (1 - \eta) / (\eta M)$ \cite{skolnik_stat_density_tapered_arrays}.

\subsection{Statistical peak-to-sidelobe level (PSL) analysis of SNB AF}
\label{sec:psl_analysis}
In the far-sidelobe region, the average PSL can be calculated from the average SL using extreme value and level crossing theory \cite{steinberg_psl_of_random_arrays, donvito_psl_of_random_arrays, psl_gumbel_distribution, ochiai_collab_beamforming, ev_psl_thesis}. 
Sufficiently far from the mainlobe, the complex fluctuation (the random sidelobes) of the AF is given by 
\begin{align}
    \label{eq:af_sidelobe_fluct}
    Z(\Delta u) &=\mathrm{AF_{th}}(\Delta u) - \mathrm{AF_{wdw}}(\Delta u).
\end{align}
Expanding the AFs from \eqref{eq:af_sidelobe_fluct} with \eqref{eq:exp_ampl_wdw_array} gives 
\begin{align}
    \label{eq:af_sidelobe_wss}
     Z(\Delta u) 
     &= \sum_m V_m e^{-j 2\pi \frac{x_m}{\lambda_{\mathrm{c}}} \Delta u},
\end{align}
where $V_m = W(x_m) - w(x_m)$
is the random variable describing the per-element weight (position) fluctuation.
The per element variance of $V_m$ is denoted by $\sigma_{V_m}^2$ and equal to
\begin{align}
    \label{eq:v_m_var}
    \sigma_{V_m}^2
    &= \frac{1}{\overline{M_{\mathrm{th}}}} \left( p_m(1 - p_m) \right) .
\end{align}
Since the occupation variables are independent, the total sidelobe variance is the sum of per-element variances $\sigma^2 = \sum_m \sigma_{V_m}^2$, consistent with \eqref{eq:af_thinned_decoupled}.
Since the occupation variables $B_m$ are independent, the autocorrelation function of the complex fluctuation \eqref{eq:af_sidelobe_wss} depends only on the lag and it is wide-sense stationary and simplifies to
\begin{align}
    \label{eq:sidelobe_autocorr}
    R_Z(\tau) 
    &=\mathbb{E} \left[ Z(\Delta u) Z^*(\Delta u - \tau) \right] \nonumber \\
    &= \sum_m \sigma_{V_m}^2 e^{-j 2\pi \frac{x_m}{\lambda_{\mathrm{c}}} \tau}.
\end{align}
 Sufficiently away from the mainlobe by the central limit theorem, the $Z(\Delta u)$ complex sidelobe fluctuation converges to a circular zero-mean complex Gaussian process.

To calculate the expected PSL, level-crossing theory \cite{rice} is employed.
The average sidelobe power of the thinned array is $\sigma^2$ as derived in \eqref{eq:af_thinned_decoupled}. 
The sidelobe envelope $|Z(\Delta u)|$ follows a Rayleigh distribution and consequently the sidelobe power follows an exponential distribution.
The expected rate at which the sidelobe power crosses a specific threshold $p$ with a positive slope (upcrossing) over a unit interval length is given by Rice's formula \cite{rice}.
Expressing the threshold power $p$ normalized to the average sidelobe power as $\rho = p / \sigma^2$, and scaling the crossing rate by a factor of $2$ to match the visible $u$ interval, the expected number of sidelobe power upcrossings is
\begin{align}
    \label{eq:n_cross_rice_rayleigh}
    \mathbb{E} \left[ N_{\mathrm{cr}}^+( \rho) \right] = 2\sqrt{\frac{\mu_2}{\pi}} \sqrt{\rho }e^{-\rho},
\end{align}
where $\mu_2$ is the second central moment of the sidelobe fluctuation from \eqref{eq:af_sidelobe_fluct}.
By the properties of the Fourier transform (and the Wiener- Khinchin theorem), the normalized second central spectral moment is given by the negative ratio of the second derivative of the autocorrelation function to the autocorrelation function, both evaluated at $\tau = 0$, as follows
\begin{align}
    \label{eq:sigma_wiener_kinchin_def}
    \mu_2 = -\frac{R_Z''(0)}{R_Z(0)}.
\end{align}
Evaluating \eqref{eq:sidelobe_autocorr} in \eqref{eq:sigma_wiener_kinchin_def} the second central moment can be written as
\begin{align}
    \label{eq:second_central_moment}
    \mu_2 
    &=\left(\frac{2\pi}{\lambda_{\mathrm{c}}} \right)^2
    \frac{\sum_m  x_m^2 \sigma_{V_m}^2}{\sum_m \sigma_{V_m}^2}.
\end{align}
For uniform spatial profile $\mathbb{E}[x_m^2] = D^2 / 12$ and $p_m = \eta$ and the element variances $\sigma_{V_m}^2$ are identical and second central moment simplifies to
$\mu_2 
\approx \pi^2 D_{\lambda}^2 / 3$.

Peak sidelobes corresponding to crossing of a high threshold by a stationary Gaussian process are rare and asymptotically independent. Consequently, they can be modeled as a Poisson point process \cite{sodin, ev_psl_thesis}. The probability of $k$ upcrossings of the normalized power $\rho$ is given by
\begin{align}
    \label{eq:poisson_proc}
    P(k, \rho) = \frac{(\mathbb{E}[N_{\mathrm{cr}}^+ (\rho)])^k}{k!} e^{-\mathbb{E}[N_{\mathrm{cr}}^+ (\rho)]}.
\end{align}
The peak-to-sidelobe power remains below a threshold $\rho$ only if no upcrossing occurs.
By setting the number of occurrences to zero ($k=0$), the cumulative distribution function (CDF) of the peak-to-sidelobe power is obtained
\begin{align}
    \label{eq:psl_cdf}
    F_{\mathrm{PSL}}(\rho) 
    &= \exp{\left(-C\sqrt{\rho }e^{-\rho} \right)}.
\end{align}
where $C = 2 \sqrt{\frac{\mu_2}{\pi}} $. The constant $C$ can be physically understood as the number of sidelobes in the visible region (or number of intependent spatial samples).
Next, the CDF of the peak power \eqref{eq:psl_cdf} is approximated by a Gumbel distribution to obtain the closed-form expressions of the mean value.
The CDF of the Gumbel distribution is given by
\begin{align}
    \label{eq:gumbel_cdf}
    F_\mathrm{G}(\rho, \mu, \beta) = \exp{\left( -e^{-\frac{\rho - \mu}{ \beta}} \right)},
\end{align}
where $\mu$ is location and $\beta$ is the scale of the distribution.
Approximating \eqref{eq:psl_cdf} as a Gumbel distribution requires linearising the natural logarithm of the argument of \eqref{eq:psl_cdf}. To achieve this, the logarithm of the argument is approximated with a first-order (linear) Taylor expansion around location $\mu$.
The natural logarithm of the argument from \eqref{eq:psl_cdf} is
\begin{align}
    \label{eq:ln_arg}
    h(\rho) = \ln C + \frac{1}{2}\ln \rho - \rho.
\end{align}
The first-order Taylor approximation with respect to $\rho$ around location $\mu$ is 
\begin{align}
    \label{eq:taylor_approx}
    h(\rho) \approx h(\mu) + h'(\mu)(\rho - \mu),
\end{align}
where $h'(\mu)
= -\frac{2\mu - 1}{2\mu}$ is the first derivative of \eqref{eq:ln_arg}.
To obtain a linear argument in \eqref{eq:taylor_approx}, the constant term $h(\mu)$ in \eqref{eq:taylor_approx} is forced to zero.
This condition yields the following transcendental equation for $\mu$ 
\begin{align}
    \label{eq:mu_transcendental}
    \mu - \frac{1}{2}\ln \mu &= \ln C.
\end{align}
The exact solution to \eqref{eq:mu_transcendental} is given by the Lambert W function \cite{lambert_w}.
For large apertures considered in this work, $\mu$ is also large and \eqref{eq:mu_transcendental} can be accurately approximated with an iterative solution.
For instance, for a uniform spatial profile 
$C \approx 2\sqrt{\frac{\mu_2}{\pi}} 
\approx 2D_{\lambda}$ and $\mu \approx \ln{2D_{\lambda}}$ 
and consequently $C$ can also be considered large. 
For large $C$ the contribution of the double logarithm can be considered negligible, yielding the first iteration solution $\mu \approx \ln{C}$. By substituting the first iteration result into \eqref{eq:mu_transcendental}, the second iteration approximation is obtained as follows
\begin{align}
    \label{eq:mu_approx}
    \mu &\approx \ln{C} + \frac{1}{2} \ln{\left( \ln{C} \right)}.
\end{align}
The second iteration provides a sufficiently accurate approximation of $\mu$ for large $C$. 
After linear approximation the argument from \eqref{eq:ln_arg} becomes
\begin{align}
    \label{eq:linearized_arg}
    h(\rho) \approx - \frac{\left( \rho - \mu \right)}{\frac{2\mu - 1}{2\mu}}
\end{align}
Reverting the logarithm in \eqref{eq:ln_arg} by taking the exponential of the linearized argument from \eqref{eq:linearized_arg} and substituting it into the exponential expression from \eqref{eq:psl_cdf}, it coincides with the Gumbel CDF from \eqref{eq:gumbel_cdf} with $\beta =\frac{2\mu }{2\mu - 1}$.
Given the approximation of the CDF of $\rho_{\mathrm{PSL}}$ as the Gumbel CDF, the expected value is given by a closed-form expression \cite{gumbel_book} as 
\begin{align}
    \mathbb{E} \left[\rho_{\mathrm{PSL}} \right]
    &= \mu + \beta \gamma
\end{align}
where $\gamma \approx 0.5772$ is the Euler-Mascheroni constant and $\beta =\frac{2\mu }{2\mu - 1}$.
The expected value of the PSL is given by reverting the normalization by the average sidelobe power $p_{\mathrm{{PSL}}} = \sigma^2 \rho_{\mathrm{PSL}} $, yielding
\begin{align}
    \label{eq:exp_psl}
 \mathbb{E}[\mathrm{PSL}] 
    = \overline{\mathrm{PSL}} 
    = \overline{\mathrm{SL}} (\mu + \beta\gamma).
\end{align}
Note that due to setting the number of crossings to 0 in \eqref{eq:n_cross_rice_rayleigh}, the derived expected PSL is an overestimate or can be considered an upper bound.

\begin{figure}[t]
    \centering
    \includegraphics[width=\linewidth]{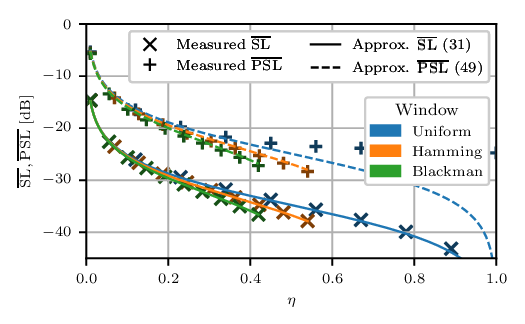}
    \caption{Comparison between measured and approximated average SL and average PSL of SNB AF as a function of $\eta$ for different spatial tapering windows. $d_{\mathrm{a}_{\lambda}} = 0.5$, $M = 3000$, 3000 Monte Carlo realizations. }
    \label{fig:sl_psl_vs_eta}
\end{figure}

\begin{figure}[t]
    \centering
    \includegraphics[width=\linewidth]{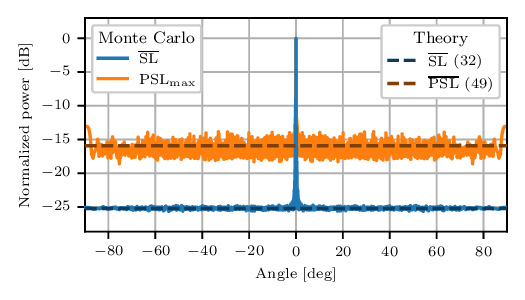}
    \caption{Comparison between the simulated and estimated SL and PSL of SNB AF for uniform spatial window $u'=0$, $d_{\mathrm{a}_{\lambda}} = 0.5$, $\eta = 0.25$, $M = 1000$ and 1000 Monte Carlo realizations.}
    \label{fig:nb_af_mc}
\end{figure}

Fig. \ref{fig:sl_psl_vs_eta} illustrates the $\overline{\mathrm{SL}}$ and $\overline{\mathrm{PSL}}$ as a function of $\eta$ for different density windows. The theoretical approximation of the peak and SL matches well with the measured results obtained by Monte-Carlo simulations.
Fig. \ref{fig:nb_af_mc} shows the SL and PSL of a SNB AF as a function of angle for a uniform spatial profile and 1000 Monte-Carlo realizations. The expected SL and PSL levels are uniform across angles and match well with levels predicted by analytical derivation.

\subsection{Approximation of statistical sidelobe suppression in SWB regime}

To provide insight into how the SWB kernel averages the random noise-like sidelobes of thinned arrays, an approximation of the SWB kernel for a uniform spectrum is introduced.
Consider that the  argument of the $\Si{}$ function in \eqref{eq:rd_norm_uni_spec} is large $\left| (D_{\lambda} + d_{\mathrm{a}_{\lambda}}) \left( \frac{B}{2 f_{\mathrm{c}}}\Delta u \pm \left( \Delta u - u_{\tau} \right) \right) \right| \gg 1$. For large aperture $D_{\lambda} \gg d_{\mathrm{a}_{\lambda}}$ the condition simplifies to $D_{\lambda} B_{\mathrm{f}} |\Delta u| \gg 1$.
For large arguments, the $\Si{}$ function converges to the sign function $\Si(x) \approx \frac{1}{2} \sgn(x)$. 
Approximating the sine integral as a sign function enables expressing the SWB kernel as a rectangular function as follows
\begin{align}
    \label{eq:rd_norm_rect}
    &\tilde{R}_{\rect{}}\left( \Delta u, -\frac{d_{\mathrm{a}}}{\lambda_{\mathrm{c}}} \left( \Delta u - u_{\tau} \right) \right)
    \overset{D_{\lambda} B_{\mathrm{f}} |\Delta u| \gg 1}{\approx} \nonumber \\
    &\qquad \qquad \qquad \overset{D_{\lambda} B_{\mathrm{f}} |\Delta u| \gg 1}{\approx}  \frac{1}{B_{\mathrm{f}} |\Delta u|} \rect{\left( \frac{\Delta u - u_{\tau}}{ B_{\mathrm{f}} |\Delta u| }\right)}.
\end{align}
Note that the total area of the kernel is conserved and equal to unity. As the kernel widens, the amplitude decreases accordingly.
Approximating the SWB kernel as a rectangular function of width $B_{\mathrm{f}}|\Delta u|$ provides an intuitive interpretation of the SWB AF as a variable-width moving average of the SNB AF.

Next, the statistical models for the expected SL and PSL from Secs. \ref{sec:sl_analysis} and \ref{sec:psl_analysis} are extended to account for SWB averaging effects on SNB AF of thinned antenna arrays.
From \eqref{eq:af_sidelobe_fluct} the AF of the thinned array can be written as a sum of a deterministic term and a complex fluctuation.
As the sidelobe power and behaviour are of interest for sufficiently large $\Delta u$, the deterministic part $\mathrm{AF_{wdw}}(\Delta u)$ in \eqref{eq:af_sidelobe_fluct} is considered negligible, leaving only the complex fluctuation (sidelobe) term $Z(\Delta u)$. For a uniform spectrum, the SWB AF sidelobes can be approximated by substituting $Z(\Delta u)$ and \eqref{eq:rd_norm_rect} into \eqref{eq:wb_af_int_u_domain}, resulting in
\begin{align}
    \label{eq:wideband_complex_fluct}
    Z_{\mathrm{SWB}}(\Delta u)
    &\approx    \frac{1}{B_{\mathrm{f}} |\Delta u|}\int_{-\infty}^{\infty} 
    Z(u_{\tau}) \rect{\left( \frac{\Delta u - u_{\tau}}{ B_{\mathrm{f}} |\Delta u| }\right)} \, du_{\tau} \nonumber \\
\end{align}
The analytical expressions for the expected SL and PSL are based on the variance of the complex fluctuation. 
The wideband sidelobe variance can be expressed as a double integral from \eqref{eq:wideband_complex_fluct} involving the autocorrelation function of the sidelobes from \eqref{eq:sidelobe_autocorr}. Next, the double integral is reduced to a single integral by evaluating the autocorrelation of a rectangular kernel, which yields a triangular function
\begin{align}
    \label{eq:swb_sigma}
    \sigma_{\mathrm{SWB}}^{2}(\Delta u) 
    &=
    \mathbb{E} 
    \left[ \left| Z_{\mathrm{SWB}}(\Delta u) \right|^2 \right]  \\
    &= \frac{1}{B_{\mathrm{f}} |\Delta u|}
    \int_{-\infty}^{\infty}
    R_{Z}(u_{\tau}) 
    \tri{\left( \frac{ u_{\tau}}{ B_{\mathrm{f}} |\Delta u| }\right)}
    \, du_{\tau}, \nonumber
\end{align}
where $\tri{(x)}$ is the triangular function.
Expanding \eqref{eq:swb_sigma} with \eqref{eq:sidelobe_autocorr} and recognizing that the Fourier transform of the triangular function is a squared sinc function gives
\begin{align}
    \label{eq:swb_sigma_simplified}
    \sigma_{\mathrm{SWB}}^{2}(\Delta u) 
    &= \sum_m \sigma_{V_m}^2 \sinc^2{\left( B_{\mathrm{f}} |\Delta u| \frac{x_m}{\lambda_{\mathrm{c}}} \right)}.
\end{align}
Note that due to the $\Delta u$ dependence of the SWB kernel, the resulting sidelobe variance and $\overline{\mathrm{SL}}$ levels also become a function of the angle.
The $\overline{\mathrm{SL}}$ of a SWB thinned array is obtained by substituting the wideband variance from \eqref{eq:swb_sigma_simplified} into \eqref{eq:exp_avg_sl}.
Calculating the expected PSL of a SWB thinned array requires updating the calculation of the second central moment from \eqref{eq:second_central_moment} with the SWB equivalent of per-element variance obtained from \eqref{eq:swb_sigma_simplified} as
\begin{align}
    \label{eq:v_m_var_swb}
    \sigma_{V_m, \mathrm{SWB}}^2 = \sigma_{V_m}^2 \sinc^2{\left( B_{\mathrm{f}} |\Delta u| \frac{x_m}{\lambda_{\mathrm{c}}} \right)}.  
\end{align}
By substituting \eqref{eq:v_m_var_swb} into \eqref{eq:second_central_moment} the SWB second central moment ($\mu_{2, \mathrm{SWB}}$) is obtained, which can then be used to calculate the SWB parameters $\mu_{\mathrm{SWB}}$ and $\beta_{\mathrm{SWB}}$. Finally, the expected PSL of the SWB thinned array is obtained by substituting the SWB parameters into \eqref{eq:exp_psl}.

\begin{figure}[tb]
    \centering
    \includegraphics[width=\linewidth]{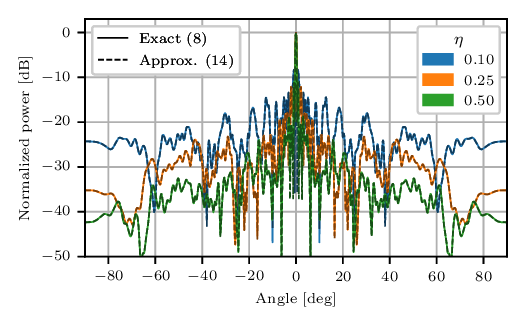}
    \caption{SWB AF of uniformly thinned array for $D_{\lambda}=100$, $d_{\mathrm{a}_{\lambda}} = 0.5$, $B_{\mathrm{f}} = 0.2$, uniform spectrum window, $u'=0$ and and selected values of the fill factor $\eta$.}
    \label{fig:wb_af_approx_vs_eta}
\end{figure}

\begin{figure}[tb]
    \centering
    \includegraphics[width=\linewidth]{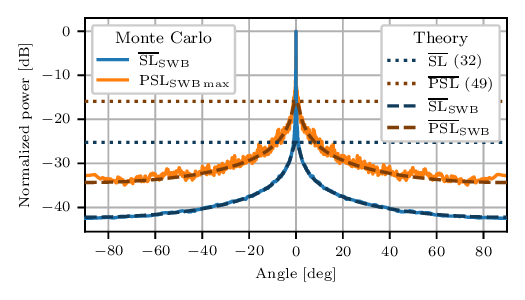}
    \caption{Comparison between the simulated and estimated SL and PSL of SWB AF for uniform spatial window $B_{\mathrm{f}} = 0.1$, $u'= 0$, $d_{\mathrm{a}_{\lambda}} = 0.5$, $\eta = 0.25$, $M = 1000$ and 1000 Monte Carlo realizations.}
    \label{fig:wb_af_mc}
\end{figure}

Fig. \ref{fig:wb_af_approx_vs_eta} shows the SWB AF of a uniformly thinned antenna array for different values of the fill factor $\eta$. Similarly, as in the SNB case, the increase in the fill factor improves the SL.
Fig. \ref{fig:wb_af_mc} illustrates the SWB AF sidelobe levels as a function of angle for $B_{\mathrm{f}} = 0.1$ and the same array parameters as in Fig.\ref{fig:nb_af_mc}.
Compared to the SNB case, the SL and PSL are reduced due to the averaging by the SWB kernel. Since the SWB kernel is angle-dependent, the resulting SL and PSL suppression also varies with angle. The analytical sidelobe level predictions agree well with Monte-Carlo simulation results.

\subsection{Approximation of SL and PSL of uniformly thinned arrays}
In this section, simple closed-form expressions of the expected SL and PSL of uniformly thinned arrays are derived.
For a uniform spatial window, the activation probability is constant, $p_m = \eta$, the per-element variance from \eqref{eq:v_m_var} is constant across $m$ and simplifies to
$\sigma_{V_m}^2 = \frac{1}{\overline{M_{\mathrm{th}}}}  \eta (1 - \eta)$.
Consequently, the sidelobe variance reduces to 
$\sigma^2 
= 1 - \eta$. 
As the per-element variance is no longer a function of the index $m$, the SWB sidelobe variance from \eqref{eq:swb_sigma_simplified} takes the form of
\begin{align}
    \label{eq:sigma_swb_uni}
    \sigma_{\mathrm{SWB}}^{2}(\Delta u) 
    &= \frac{\sigma^2}{M} \sum_m  \sinc^2{\left( B_{\mathrm{f}} |\Delta u| \frac{x_m}{\lambda_{\mathrm{c}}} \right)}.
\end{align}
Next, to arrive at a closed-form expression, the sum in \eqref{eq:sigma_swb_uni} is approximated as an integral based on the Riemann sum approximation $\sum_m f(x_m) \approx \frac{1}{\Delta x} \int f(x)\, dx$, similarly to \eqref{eq:cont_rd_approx}. The approximation is accurate given that the approximated function is smooth and slowly varying across $m$. For discretized uniformly thinned arrays the condition is $\Delta x_m B_{\mathrm{f}} |\Delta u| \ll 1$, where $\Delta x_m = d_{\mathrm{a}_{\lambda}}$. Given the maximum value of $|\Delta u|$ is 2 the approximation condition simplifies to $B_{\mathrm{f}} \ll 1 / (2 d_{\mathrm{a}_{\lambda}})$. The continuous approximation of \eqref{eq:sigma_swb_uni} can be expressed as
\begin{align}
    \label{eq:sigma_swb_uni_int}
    \sigma_{\mathrm{SWB}}^{2}(\Delta u) 
    & \overset{B_{\mathrm{f}} \ll 1 / (2 d_{\mathrm{a}_{\lambda}})}{\approx} 
    \frac{\sigma^2}{D_{\lambda}} \int_{-D_{\lambda}/2}^{D_{\lambda}/2}  \sinc^2{\left( B_{\mathrm{f}} |\Delta u| x_{\lambda} \right)} \, dx_{\lambda} \nonumber \\
    &= \sigma^2 
    \frac{2}{\nu} 
    \left( 
    \Si{\left( \nu \right)} -  
    \frac{2 \sin^2{(\pi \nu} / 2 ) }
    {\pi^2 \nu}
    \right),
\end{align}
where $\nu = D_{\lambda} B_{\mathrm{f}} |\Delta u|$ is introduced here for compactness of the final expression.
Under the large $\nu \gg 1$ assumption introduced earlier in \eqref{eq:rd_norm_rect} leading to this derivation, the Si function converges to $1/2$ and the second term fraction can be considered negligible, yielding the approximation of the SWB variance
\begin{align}
    \label{eq:sigma_swb_uniform_large_approx}
    \sigma_{\mathrm{SWB}}^{2}(\Delta u)  
    &\overset{\nu \gg 1}{\approx}
    \frac{\sigma^2}{D_{\lambda} B_{\mathrm{f}} |\Delta u| }. 
\end{align}
Substituting the approximation of the SWB variance from \eqref{eq:sigma_swb_uniform_large_approx} into the expected SL expression from \eqref{eq:exp_avg_sl_even_nonneg} gives
\begin{align}
    \label{eq:sl_swb_uni_approx}
    \overline{\mathrm{SL}}_{\mathrm{uni, SWB}} 
    &= \frac{1 - \eta}{\eta M} \frac{1}{D_{\lambda} B_{\mathrm{f}} |\Delta u| }
    =  \frac{\overline{\mathrm{SL}}_{\mathrm{uni}}}{D_{\lambda} B_{\mathrm{f}} |\Delta u| }.
\end{align}
The SWB regime introduces scaling by the factor $D_{\lambda} B_{\mathrm{f}} |\Delta u|$ as compared to the SNB SL expression.

Obtaining a closed-form expression for the expected PSL of a uniformly thinned array 
requires approximating the discrete sums in \eqref{eq:second_central_moment} by integrals.
After expanding \eqref{eq:second_central_moment} with the per-element variance from \eqref{eq:v_m_var_swb}, the discrete sums are approximated as integrals following the same approach and criteria as in \eqref{eq:sigma_swb_uni_int}.
The continuous approximation of the second central moment for a uniform spatial profile is
\begin{align}
    \label{eq:second_central_moment_swb_cont_approx}
    &\mu_{2, \mathrm{SWB, uni}} 
    \overset{B_{\mathrm{f}} \ll 1 / (2 d_{\mathrm{a}_{\lambda}})}{\approx} 
    \left(2\pi \right)^2
    \frac{\int_{-D_{\lambda}/2}^{D_{\lambda}/2}  x_{\lambda}^2  \sinc^2{\left( B_{\mathrm{f}} |\Delta u| x_{\lambda} \right)} \, dx_{\lambda}}
    {\int_{-D_{\lambda}/2}^{D_{\lambda}/2}  \sinc^2{\left( B_{\mathrm{f}} |\Delta u| x_{\lambda} \right)} \, dx_{\lambda} } \nonumber \\
    &\qquad = \left(\frac{2}{B_{\mathrm{f}} |\Delta u|} \right)^2
    \frac{
    \left( \frac{D_{\lambda}}{2} - \frac{\sin{(\pi\nu )}}{2 \pi B_{\mathrm{f}} |\Delta u|} \right)
    }
    {\frac{2}{B_{\mathrm{f}} |\Delta u| }  \left( 
    \Si{\left(\nu \right)} -  
    \frac{2 \sin^2{ \left(\pi\nu / 2 \right)} }
    {\pi^2\nu}
    \right) }
\end{align}
Similarily as in \eqref{eq:sigma_swb_uniform_large_approx} for large $\nu \gg 1$  \eqref{eq:second_central_moment_swb_cont_approx} simplifies to
\begin{align}
    \mu_{2, \mathrm{SWB, uni}} 
    &\overset{\nu \gg 1}{\approx}
    \frac{2D_{\lambda}}{B_{\mathrm{f}} |\Delta u|}.
\end{align}
Given the approximated second central moment, the SWB parameter $\mu_{\mathrm{SWB}}$ from \eqref{eq:mu_approx} can be written as
\begin{align}
    \mu_\mathrm{SWB, uni} = \ln{\left( 2 \sqrt{\frac{2D_{\lambda}}{\pi B_{\mathrm{f}} |\Delta u|}} \right)} + \frac{1}{2} \ln{\left( \ln{\left( 2 \sqrt{\frac{2D_{\lambda}}{\pi B_{\mathrm{f}} |\Delta u|}} \right)} \right)}
\end{align}
Substituting the approximated parameters into \eqref{eq:exp_psl}, the closed-form expression of the expected PSL of a uniformly thinned array is obtained as follows
\begin{align}
    \overline{\mathrm{PSL}}_\mathrm{SWB, uni} 
    = \overline{\mathrm{SL}}_\mathrm{SWB, uni} 
    \left(\mu_\mathrm{SWB, uni}  + \beta_\mathrm{SWB, uni}  \gamma \right),
\end{align}
where $\beta_\mathrm{SWB, uni} = \frac{2\mu_\mathrm{SWB, uni}}{2\mu_\mathrm{SWB, uni} - 1}$.

\begin{figure}[tb]
    \centering
    \includegraphics[width=\linewidth]{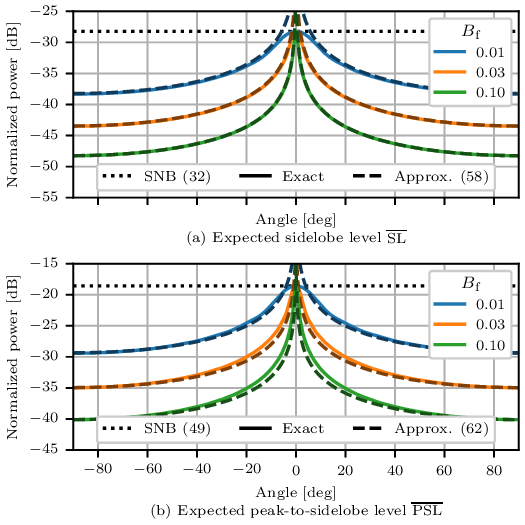}
    \caption{Comparison between the exact analytical expressions and approximations of the expected SL and PSL in SWB AF for uniform spectrum and uniform spatial profile $u'=0$, $D_{\lambda} = 1000$ and varying $B_{\mathrm{f}}$.}
    \label{fig:wb_sl_psl_vs_prod}
\end{figure}

Fig. \ref{fig:wb_sl_psl_vs_prod} illustrates the accuracy of the approximations of expected SL and PSL in SWB AF obtained under the large $\nu \gg 1$ assumption. The approximation accuracy is degraded near the main lobe where $\nu$ is small. An increase in fractional bandwidth $B_{\mathrm{f}}$ or aperture $D_{\lambda}$ improves the SL and PSL sidelobe suppression. Both the expected SL and PSL in the SWB regime are approximately inversely proportional to the $D_{\lambda} B_{\mathrm{f}} |\Delta u|$ product, resulting in more effective suppression for large $|\Delta u|$ - away from the mainlobe.

\section{Conclusion}
This work extends the concept of SWB AF to thinned arrays.
The SWB AF approximation as a spatially variant convolution is shown to remain valid and accurate, offering insight into how the large bandwidth-aperture product modifies the SNB AF and its sidelobes. 
In thinned arrays, the sidelobes of the SNB AF away from the mainlobe can be modeled as white Gaussian noise, making suppression by the SWB kernel particularly effective.
The analytical formulas for expected SL and PSL levels of SNB AF are extended to the SWB regime, allowing accurate characterization for arbitrary element density profiles.
Finally, closed-form expressions for the SL and PSL levels of SWB AF of the uniformly thinned arrays with uniform spectrum are derived. The results show that both the expected SL and PSL levels are inversely proportional to the bandwidth-aperture product and angle difference.
Due to the dependence of the SL and PSL suppression on the angle difference, it becomes increasingly effective away from the mainlobe.

\bibliographystyle{IEEEtran}
\bibliography{IEEEabrv.bib, biblio.bib}

\end{document}